\documentclass[aps,prd,superscriptaddress,nofootinbib,amsfonts,amssymb,amsmath,notitlepage]{revtex4-1}

\usepackage[dvipdfmx]{graphicx}
\usepackage{amsmath}
\usepackage{amssymb} \allowdisplaybreaks[4]
\usepackage{afterpage,dblfloatfix}
\usepackage{color}
\usepackage{xcolor}
\usepackage[colorlinks=true, allcolors=blue!75!black]{hyperref}
\usepackage{makecell}
\usepackage{multirow}
\usepackage{slashed}
\usepackage{wrapfig}
\usepackage{placeins}
\usepackage{bm}
\usepackage[nameinlink]{cleveref}
\usepackage[caption=false]{subfig}

\allowdisplaybreaks[3]

\begin{document}

\title{
Axionlike particle-assisted supercooling chiral phase transition in QCD: \\
Identifying Coleman-Weinberg type-chiral phase transition in QCD-like scenarios 
}

\author{Zheng-liang Jiang}\thanks{{\tt jiangzl25@mails.jlu.edu.cn}}
\affiliation{Center for Theoretical Physics and College of Physics, Jilin University, Changchun, 130012,
China}

\author{Yuepeng Guan}\thanks{{\tt guanyp22@mails.jlu.edu.cn}}
\affiliation{Center for Theoretical Physics and College of Physics, Jilin University, Changchun, 130012,
China}

\author{Mamiya Kawaguchi}\thanks{{\tt mamiya@aust.edu.cn}} 
      \affiliation{ 
Center for Fundamental Physics, School of Mechanics and Physics,
Anhui University of Science and Technology, Huainan, Anhui 232001, People’s Republic of China
}

\author{Shinya Matsuzaki}\thanks{{\tt synya@jlu.edu.cn}}
\affiliation{Center for Theoretical Physics and College of Physics, Jilin University, Changchun, 130012, China}%

\author{Akio Tomiya}\thanks{{\tt akio@yukawa.kyoto-u.ac.jp}} 
\affiliation{Department of Information and Mathematical Sciences, Tokyo Woman’s Christian  University, Tokyo 167-8585, Japan} 
\affiliation{RIKEN Center for Computational Science, Kobe 650-0047, Japan} 
%RIKEN BNL Research center, Brookhaven National Laboratory, Upton, NY, 11973, USA 

\author{He-Xu Zhang}\thanks{{\tt zhanghexu@ucas.ac.cn}}
\affiliation{School of Nuclear Science and Technology, University of Chinese Academy of Sciences, Beijing 100049, China}

\begin{abstract} 

We propose a new scenario to realize the Coleman-Weinberg (CW) type chiral phase transition in the QCD thermal history. This scenario predicts a heavy axionlike particle (ALP) with mass $\sim$ 5 MeV, consistently with the current experimental and cosmological bounds.  
The chiral phase transition is evaluated by monitoring ordinary QCD setup in a view of 
a two-flavor Nambu-Jona-Lasinio model including a simplified meson fluctuation contribution. The present work thus can open a new window to search for 
the ALP associated with the QCD phase transition epoch of the thermal history. 
The new QCD cosmological scenario potentially predicts rich epochs around the QCD scale: a mini-inflation; a nonperturbative preheating and/or reheating, which can provide characteristic gravitational wave and primordial black hole productions. 
This proposal is based on a generic classification of the order of the chiral phase transition at the level of the mean field approximation 
in view of the scale violation classes: the soft-scale breaking term and 
the CW-type scale anomaly term, in or off the medium with or without chemical potentials. 
On this theoretical ground, we also revisit existing scenarios which undergo the supercooling chiral phase transition, such as nearly scale-invariant QCD and QCD with a large baryon chemical potential.

\end{abstract} 
\maketitle

\section{Introduction %and summary
} 

First order phase transitions or supercooling phase transitions currently involve broader interests over theoretical particle physics and cosmology~\cite{Mazumdar:2018dfl}. 
A particular motivation can be seen in cosmological applications to the thermal history of the Universe: the first-order/supercooling phase transition serves as productions for stochastic gravitational wave (GW) background and primordial black hole (PBH) formation~\cite{Khlopov:2008qy,Jung:2021mku,Kodama:1982sf,Hall:1989hr,Kusenko:2020pcg,Hashino:2021qoq,Hashino:2022tcs,He:2022amv,Liu:2021svg,Kawana:2022olo,Lewicki:2023ioy,Gouttenoire:2023naa,Gouttenoire:2023pxh,Salvio:2023blb,Conaci:2024tlc,Banerjee:2023brn,Banerjee:2023vst,Banerjee:2023qya,Lewicki:2024ghw,Flores:2024lng,Cai:2024nln,Florentino:2024kkf,Goncalves:2024vkj,Banerjee:2024fam,Shao:2024dxt,Banerjee:2024cwv,Wu:2024lrp,Hashino:2025fse,Murai:2025hse,Franciolini:2025ztf,Kierkla:2025vwp,Kanemura:2024pae} as nonparticle dark matter candidate~\cite{Belotsky:2018wph,Carr:2020gox,Baldes:2023rqv,Balaji:2024rvo,Jinno:2023vnr,Zhang:2025kbu,Kanemura:2024pae}. 
It has also been argued that the supercooling electroweak or QCD phase transition can provide a possible initial condition for slow-roll inflation~\cite{Zhang:2023acu,Liu:2024xrh} and subsequent particle production~\cite{Wang:2022ygk,Wang:2024tda}.

One typical modeling to access those cosmological applications would be a scale-symmetry phase transition via the Coleman-Weinberg (CW) mechanism~\cite{Coleman:1973jx}, assuming an appropriate presence of scalar field (presumably allowing coupled to the Standard Model (SM) plasma). 
All dimensionful parameters are absent at tree level, but quantum loop corrections induce the scale anomaly and generate the CW potential leading to spontaneous scale-symmetry breaking. 
Since the CW potential merely includes a logarithmic field dependence, it exponentially suppresses the tunneling rate at low temperatures. This intrinsic feature naturally induces strong enough supercooling, therefore, is particularly relevant for PBH formation~\cite{Gouttenoire:2023naa,Gouttenoire:2023pxh,Baldes:2023rqv,Conaci:2024tlc,Arteaga:2024vde,Salvio:2023ynn,Salvio:2023blb,Banerjee:2024cwv,Kierkla:2025vwp}.

Very recently, the importance of what is called the ``soft scale-breaking" mass term has also been emphasized in addressing the CW-type supercooling phase transition~\cite{Zhang:2025kbu}. 
This soft-scale breaking term can dramatically alter the prediction for PBH formation, due to violation of the exponential nucleation approximation. 
However, it still leaves the successful stochastic GW background predictions essentially as they are in the case of the pure CW scenario~\cite{Jaeckel:2016jlh,Marzola:2017jzl,Iso:2017uuu,Ghorbani:2017lyk,Baldes:2018emh,Prokopec:2018tnq,Brdar:2018num,Marzo:2018nov,DelleRose:2019pgi,Ghoshal:2020vud,Levi:2022bzt,Kierkla:2022odc,Salvio:2023qgb,Zhang:2024vpp}. 
No matter how small such a deformation is, the soft-scale breaking 
is still active even at a low enough temperature -- much lower than the critical temperature for the thermal phase transition ($T_c$), but as low as the critical temperature for the bubble nucleation ($T_n$).  
 Thus this deformation possibility has provided a new testable link between PBH dark matter and GW signatures in the CW-type scenario~\cite{Zhang:2025kbu}.

The potential form of the deformed CW potential addressed in~\cite{Zhang:2025kbu} essentially goes like 
\begin{align} 
V(M: T \ll T_c) \sim  m_0^2 M^2 + c_1 \cdot M^4 [ \ln (M/v) + c_2]
\,.\label{essential}
\end{align}
Here $M$ is a CW scalar field; the $m_0^2$ term corresponds to what is called the soft-scale breaking term; $v$ denotes the vacuum expectation value of $M$; $c_{1,2}$ are 
parameters to be determined at one-loop level given a model. 
The potential barrier around the origin can be generated when $m_0^2>0$ with $c_1>0$, 
much like in the case of the CW-type thermal phase transition. 
As long as $m_0^2$ is small enough, this phase transition will be supercooled due to 
the approximate scale invariance around the origin of the potential.

Actually, the potential form Eq.(\ref{essential}) is also the one that arises in low-energy chiral effective models of QCD and QCD-like theories; e.g., a Nambu-Jona-Lasinio (NJL) description in the mean-field approximation (MFA) with the bosonization of chiral quark bilinear: $M \sim \bar{q}q$ and quark meson models with quark and meson fluctuation corrections included~\cite{Mocsy:2004ab,Khan:2011wai,Zacchi:2017ahv}. 
In there, the logarithmic term $\sim M^4 \ln M$ is generated as a consequence of 
the quantum scale anomaly, which is also related to the renormalization group evolution of the (dynamically induced) quartic coupling for $M$. 
In medium with, e.g., finite baryon chemical potential $\mu_B$, this term would get corrections depending on $\mu_B$.  
The soft-scale breaking term $\sim m_0^2 M^2$, at $T \sim T_n \ll T_c$, would be governed by the vacuum contribution, which induces $m_0^2 < 0$ as the spontaneous chiral symmetry breaking, and also Debye screening mass in the medium, $\sim \mu_B^2 M^2$, which 
drives $m_0^2$ to be positive. 
%Furthermore, beyond the SM contribution, if coupled to the QCD chiral phase transition, %might also come into play as the soft-scale breaking term. 
Thus, the chiral phase transition in QCD-like theories 
can also be argued in terms of the same class of scale violation, soft-scale breaking $\sim M^2$, and the quantum scale anomaly $\sim M^4 \ln M$, 
as discussed in~\cite{Zhang:2025kbu}.

In QCD of the SM, which we shall call ordinary QCD, 
the chiral phase transition is a crossover around the vanishing baryon chemical 
potential~\cite{Aoki:2006we,Borsanyi:2020fev,Steinbrecher:2018phh}. 
Thereby, a first-order and supercooling phase transition would be unlikely to take place 
within the ordinary QCD framework. 
This feature is also reflected in the NJL-MF description (see, e.g., reviews~\cite{Hatsuda:1994pi}) and quark-meson models (e.g., \cite{Mocsy:2004ab,Khan:2011wai,Zacchi:2017ahv} for the case including both quark and meson fluctuations). 
As will be more explicitized later, 
the deviation from the CW-type phase transition is tightly linked with the fact that 
ordinary QCD generates a negatively too large soft-scale breaking mass ($m_0^2 <0$ in Eq.(\ref{essential})).

In the literature~\cite{Shao:2024dxt,Shao:2024ygm}, the possibility of realizing the supercooling from the QCD phase transition has been explored based on a couple of low-energy effective models. 
There, it has been clarified that with a large quark chemical potential of ${\cal O}( 400\,{\rm MeV})$, a small phase transition rate $(\beta/H)$ can be realized and produce nano hertz GWs. 
This, in terms of soft-scale breaking, occurs due to 
not merely the creation of the Fermi surface, but the Debye mass $\sim \mu_B^2 M^2$ that is positive and still large enough even at $T \sim T_n$.

Another Debye mass effect as a positive soft-scale breaking term has also been discussed 
in the recent literature~\cite{Zhang:2025kbu}. 
In the reference, in a context different from ordinary QCD, 
the chiral (axial) chemical potential $\mu_5$ in a model of scale-invariant dark QCD 
does the job along with mixing with an axial charged axionlike particle (ALP).

Thus it would be suspected that the scale violation classification, as briefly illustrated above, can give a unified identifier of the chiral phase transition and a  discriminator for the first-order. 
More remarkably, one could identify the CW-type chiral phase transition even in QCD-like theories. This might be required to have new physics coupled to QCD or QCD-like theories. 
Pursing this possibility would pave a new way toward hints for new physics coupled or related to the QCD chiral phase transition in light of GW productions and PBH formation.

In this paper, we first recap and clarify how the chiral phase transition in 
QCD-like theories can be CW-type with Beyond the SM. 
The key phenomenon is the cancellation of the soft-scale breaking terms around $T=T_c$. 
We discuss an explicit dynamics to realize this key condition, which includes a well-known feature of many-flavor (walking) dark QCD and the recently addressed chiral phase transition with a large $\mu_B$ $\sim 1$ GeV in QCD~\cite{Shao:2024dxt,Shao:2024ygm}, where for the latter, the Affleck-Dine baryogenesis~\cite{Affleck:1984fy,Linde:1985gh,Dine:2003ax} is assumed to leave a large amount of $\mu_B$ in the QCD phase transition epoch.

The CW-type supercooling phase transition can be realized even in QCD-like theories, 
where the phase transition essentially 
takes the potential form in Eq.(\ref{essential}). 
This is possible as long as the quantum scale anomaly takes a four- or three-dimensional form. 
The strongly magnetized medium is, therefore, still nontrivial 
because the four-dimensional system is to be reduced to the two-dimensional one characterized as $SO(2) \times SO(1,1)$ symmetry. 
This case will be disregarded in the present study.

We employ a two-flavor NJL model in the MFA, with monitoring the QCD case, and analyzing the thermal chiral phase transition in the presence of the chiral chemical potential $\mu_5$, nonzero strong CP phase $\theta$, in addition to the baryon chemical potential $\mu_B$. 
The scale violation classification is also operative in the medium characterized by other chemical potentials, such as the isospin chemical potential ($\mu_I$), as long as 
the extreme condition is supplied by external gauging of the global chiral symmetry.

The three-flavor QCD case would have another driving force of the trilinear term $M^3$, which can also generate 
a potential barrier, associated with the $U(1)$ axial anomaly and the QCD instanton~\cite{Kobayashi:1970ji,Kobayashi:1971qz,tHooft:1976snw}. 
This cubic term, however, is not essential for the spontaneous breaking of 
the chiral symmetry -- which does not destabilize the symmetric phase at $M=0$ and acts as a higher order correction, when the negative $M^2$ term is present -- 
and does not get thermal corrections at the quark-and/or meson one-loop level. 
Therefore, this cubic term would be subdominant when the CW-type chiral phase transition is realized and the supercooling is processed.

Here we highlight generic features that will be clarified in the present paper:

\begin{itemize}

\item 
In NJL-like models or quark meson-like models, the several types of sources for ``soft"-scale violation, 
up to temperature $T$, are introduced as \\ 
\\
i) charge or matter density by gauging global currents (in part); 
e.g., $\mu_B$, $\mu_5$, the Polyakov loop (as an imaginary chemical potential), 
\\ 
ii) current quark mass terms $m$ and $m e^{i \theta}$; 
\\ 
iii) dynamical chiral symmetry breaking via four-fermion interactions and/or negative mass square for the chiral-order parameter-field $M$.

 \item At high enough $T$, the thermal correction destructively cancels the CW-type scale anomaly of the form $\sim M^4 \log M$ for the quark loop at vacuum, while for meson loops, the thermal correction is constructive, so the CW-type scale anomaly survives. 
 %This is how in the NJL-MF description, the chiral phase transition is triggered simply by the %Debye mass generated for $M$ and thus goes like the second-order type, while inclusion of meson $fluctuation potentially tends to generate the first order of the CW-type. 

   \item 
   If the scale symmetry at the vacuum is only logarithmically broken, like CW-type models, 
   the high-$T$ expansion is insufficient to see the thermal phase transition, which is because the transition can take place at relatively lower $T$ in comparison with the case where a soft scale-breaking, negative mass-squared term $(\sim - M^2)$ is present. 
   In that case, even the fermion-loop induced CW-type scale anomaly is not perfectly canceled enough to interfere the creation of the potential barrier around $M=0$. 
%   Thus this case undergoes a first-order phase transition, just like the CW-type models. 

  \item 
Compared with the CW-type model, NJL-MF theories induce spontaneous chiral symmetry breaking, which reflects the negative mass-squared term for $M$. This deepens the curvature around $M=0$, so that the transition temperature becomes high enough that the high-$T$ expansion works well and 
cancels the net logarithmic term to wash out the barrier wall. 
Thus, this causes the second-order, or crossover phase transition. 
%    This is why the first order is not realized along the $T$-axis in QCD-like theories. 
%    This also implies a key understanding of how the electroweak phase transition becomes a %crossover in the Standard Model with the negative soft-scale breaking term $\sim - H^2$. 

\item 
With nonzero $\mu_B$ at $T=0$, in NJL-MF models, the CW-type scale anomaly of the form $\sim M^4 \log M$ is also canceled by the finite density ($\mu_B$) correction, in a way similar to 
the high $T$ correction case. 
Still, the finite $\mu_B$ generates not only the Debye mass for $M$, but also a potential barrier due to the creation of the Fermi surface which acts as a repulsive force. Therefore, at $T=0$, the chiral phase transition is of first order, which is not generically of the CW type form. 
However, at finite $T$, it is possible to have the CW-type phase transition with an appropriate $\mu_B$ by controlling the net soft-scale breaking mass term.

\end{itemize}

Passing through the clarification of the scale violation classes as above, 
we propose a new QCD cosmology scenario: the ALP-assisted CW-type chiral supercooling. 
The desired cancellation of the soft-scale breaking terms is achieved 
around $T=T_c$ by the portal coupling of the QCD-chiral order-parameter to an ALP field, which contributes significantly along with the chiral chemical potential, $\mu_5$, induced by the vacuum transition of the QCD sphaleron. 
It turns out that this scenario predicts rich and new QCD cosmology, which possibly includes a mini-inflation; preheating; reheating, and GW and/PBH production. 
As $T$ cools below as much as or less than $T_n$, where $\mu_5$ goes away due to the QCD sphaleron decoupling, in the chiral broken/confinement phase,  
we observe a massive ALP like a light axion, and the mass is constrained to be $\sim 5$ MeV, so as to survive the current experimental and cosmological bounds on the photon coupling $g_{a\gamma\gamma}$.

This paper is structured as follows: 
in Sec.~\ref{sec2}, we present the preliminary setup of the two-flavor NJL model 
and derive the thermodynamic potential in the MFA with nonzero $\mu_5, \mu_B, \theta$ at finite $T$. 
In Sec.~\ref{sec3-main},  
we clarify the generic properties of the thermal chiral phase transition at $\mu_B=0$
in comparison with the CW-type phase transition, particularly paying attention to 
the scale violation classes, without $\mu_B$ (Sec.~\ref{sec3}) and with $\mu_B$ (Sec.~\ref{sec4}).  
Then, in Sec.~\ref{ALP}, we propose the ALP-assisted CW-type chiral supercooling and 
present the phenomenological constraints on the predicted ALP and discuss its experimental and cosmological constraints. 
The conclusion of this paper is provided in Sec.~\ref{concl}.

\section{Two-flavor mean-field NJL monitor} \label{sec2}

We employ a two-flavor NJL model 
with the lightest quark doublet $q=\left(u,d\right)^{T}$ 
and focus on the scale violation classes as has briefly been addressed in the Introduction. 
The Lagrangian is given as 
\begin{align}\label{eq:1}
\mathcal{L} &=\bar{q}\left(i\gamma_{\mu}\partial^{\mu}-m+\frac{1}{3}\mu_{B}\gamma^{0}+\mu_{5}\gamma^{0}\gamma^{5}\right)q 
\notag\\ 
& +\frac{g_{s}}{2}\sum^{3}_{a=0}\left[\left(\bar{q}\tau_{a}q\right)^{2}+\left(\bar{q}i\gamma_{5}\tau_{a}q\right)^{2}\right]
+g_{d}\left(e^{i\theta}\text{det}\left[\bar{q}\left(1+\gamma_{5}\right)q\right]+e^{-i\theta}\text{det}\left[\bar{q}\left(1-\gamma_{5}\right)q\right]\right)
\,, 
\end{align}
where we have taken the isospin symmetric limit: $m_u=m_d=m$, and have also introduced 
the determinant term related to the $U(1)_A$ anomaly in the underlying QCD; 
$\tau_a$ denotes the Pauli matrices with $\tau_0=1_{2\times 2}$. 
The current quark mass is defined to be real and positive, so that the strong CP phase $\theta$ coupled to the $U(1)$ axial-anomalous determinant term has already involved the quark mass phase, i.e., $\theta \equiv \bar{\theta}$. 
Regarding the scale violation classes at this point, as listed in the Introduction, 
we see that   
\begin{itemize} 
\item[i)] 
the baryon chemical potential $\mu_B$, the chiral chemical potential $\mu_5$, arising 
via gauging the $U(1)$ vectorial and axial global symmetries, respectively;   

\item[ii)] the current quark mass $m$; 

\item[iii)] the four-fermion interactions along with the couplings $g_s$ and $g_d$, which are the essential driving force to trigger dynamical chiral symmetry breaking. 

\end{itemize}

It is convenient to transform quark fields so as to remove $\theta$ from the $g_d$-  determinant term, by a U(1) axial rotation
%\begin{equation}\label{eq:2}
$
q\to q^{\prime} =  e^{-i\gamma_{5}\frac{\theta}{4}}q  
$. 
%\end{equation}
The Lagrangian in Eq.(\ref{eq:1}) is then transformed as  
\begin{align} 
\label{eq:3}
\mathcal{L} \to \mathcal{L}^{\prime} 
& 
=\bar{q}^{\prime}\left(i\gamma_{\mu}\partial^{\mu}-
m\left[\text{cos}\frac{\theta}{2}+i\gamma_{5}\text{sin}\frac{\theta}{2}\right]
+\frac{1}{3}\mu_{B}\gamma^{0}+\mu_{5}\gamma^{0}\gamma^{5}\right)q^{\prime} 
\notag\\ 
& 
+\frac{g_{s}}{2}\sum^{3}_{a=0}\left[\left(\bar{q}^{\prime}\tau_{a}q^{\prime}\right)^{2}+\left(\bar{q}^{\prime}i\gamma_{5}\tau_{a} q^{\prime}\right)^{2}\right]+g_{d}\left(\text{det}\left[\bar{q}^{\prime}\left(1+\gamma_{5}\right)\bar{q}'\right]+\text{h.c.}\right)
\,. 
\end{align}

We work in the MFA and take into account the mean fields only for the $SU(2)$ isospin singlets~\footnote{This assumption is ensured by the absence of the isospin chemical potential, and our current discussions can also be straightforwardly extended by including the isospin breaking.}.   
Then the target mean fields are those for the following bilinears: 
$\left(\bar{q}^{\prime}_{i}q^{\prime}_{i}\right)$ and $\left(\bar{q}^{\prime}_{i}i\gamma_{5}q^{\prime}_{i}\right)$. 
Those are expanded around the mean fields $S^{\prime}=\langle\bar{q}^{\prime}_{i}q^{\prime}_{i}\rangle$ and $P^{\prime}=\langle\bar{q}^{\prime}_{i}i\gamma_{5}q^{\prime}_{i}\rangle$ as  
$\bar{q}^{\prime}_{i}q^{\prime}_{i}=S^{\prime}+\left(:\bar{q}^{\prime}_{i}q^{\prime}_{i}:\right)$ and $\langle\bar{q}^{\prime}_{i}i\gamma_{5}q^{\prime}_{i}\rangle=P^{\prime}+\left(:\bar{q}^{\prime}_{i}i\gamma_{5}q^{\prime}_{i}:\right)$~\footnote{
The U(1) axial rotation relates the scalar and pseudoscalar bilinears between the original- and prime-base scalar and pseudoscalar bilinears (for each quark flavor $i$) as
\begin{align} %\label{eq:5}
\left(\bar{q}_{i}q_{i}\right)&=\left(\bar{q}^{\prime}_{i}q^{\prime}_{i}\right)\text{cos}\frac{\theta}{2}+\left(\bar{q}^{\prime}_{i}i\gamma_{5}q^{\prime}_{i}\right)\text{sin}\frac{\theta}{2}\,, \notag\\
\left(\bar{q}_{i}i\gamma_{5}q_{i}\right)&=-\left(\bar{q}^{\prime}_{i}q^{\prime}_{i}\right)\text{sin}\frac{\theta}{2}+\left(\bar{q}^{\prime}_{i}i\gamma_{5}q^{\prime}_{i}\right)\text{cos}\frac{\theta}{2}
\,. \notag  
\end{align}
}. 
Here the terms sandwiched by ":" stand for the normal order product, which means $\langle:\mathcal{O}:\rangle=0$ for $\mathcal{O}=S^{\prime},P^{\prime}$, hence ${\cal L} = {\cal L}|_{\rm MF-only} + (: {\cal L}:)$.  
\if{ 
Then the interaction terms are replaced as 
\begin{align}\label{eq:6}
\left(\bar{q}^{\prime}_{i}q^{\prime}_{i}\right)^{2}&\to 4S^{\prime}\sum_{i}\left(\bar{q}^{\prime}_{i}q^{\prime}_{i}\right)-4S^{\prime 2}\nonumber\\
\left(\bar{q}^{\prime}_{i}i\gamma_{5}q^{\prime}_{i}\right)^{2}&\to 4P^{\prime}\sum_{i}\left(\bar{q}^{\prime}_{i}i\gamma_{5}q^{\prime}_{i}\right)-4P^{\prime 2}\nonumber\\
\text{det}\left[\bar{q}^{\prime}_{i}\left(1+\gamma_{5}\right)q^{\prime}_{i}\right]+\text{h.c.}&\to
2S^{\prime}\sum_{i}\left(\bar{q}^{\prime}_{i}q^{\prime}_{i}\right)-2P^{\prime}\sum_{i}\left(\bar{q}^{\prime}_{i}i\gamma_{5}q^{\prime}_{i}\right)-2\left(S^{\prime 2}-P^{\prime 2}\right)
\end{align}
}\fi
Thus, in the MFA, the Lagrangian in Eq.(\ref{eq:3}) takes the form
\begin{equation}\label{eq:7}
{\cal L}_{\rm MF-only}\equiv  
\mathcal{L}_{\text{MFA}}=\sum_{i}\bar{q}^{\prime}_{i}\left(i\gamma_{\mu}\partial^{\mu}-(\alpha+i\gamma_{5}\beta) +\frac{1}{3}\mu_{B}\gamma^{0}+\mu_{5}\gamma^{0}\gamma^{5}\right)q^{\prime}_{i}-2g_{s}\left(S^{\prime 2}+P^{\prime 2}\right)-2g_{d}\left(S^{\prime 2}-P^{\prime 2}\right)
\,, 
\end{equation}
where
\begin{align}\label{eq:8}
%\mathcal{M}&=\alpha+i\gamma_{5}\beta \nonumber\\
\alpha&=m\medspace\text{cos}\frac{\theta}{2}-2\left(g_{s}+g_{d}\right)S^{\prime}
\,, \nonumber\\
\beta&=m\medspace\text{sin}\frac{\theta}{2}-2\left(g_{s}-g_{d}\right)P^{\prime}
\,. 
\end{align}

By integrating out quark fields and applying the imaginary time formalism, 
we get the thermodynamic potential $\Omega$,   
\begin{equation}\label{eq:9}
\Omega=2 g_{s}\left(S^{\prime 2}+P^{\prime 2}\right)+2g_{d}\left(S^{\prime 2}-P^{\prime 2}\right)
-N_{c}N_{f}\sum_{s=\pm1}\int\frac{d^{3}p}{(2\pi)^{3}}\left[E_{s}+T\text{ln}\left(1+e^{-\frac{(E_{s}+\frac{1}{3}\mu_{B})}{T}}\right)+T\text{ln}\left(1+e^{-\frac{(E_{s}-\frac{1}{3}\mu_{B})}{T}}\right)\right]
\,, 
\end{equation}
where $N_{f}=2$, $N_{c}=3$ for the QCD monitor, and 
\begin{align}
   E_{s}=\sqrt{M^{2}+(|p|-s\mu_{5})^{2}} 
   \,, \qquad 
   M^{2}=\alpha^{2}+\beta^{2} 
   \,. \label{Es-M}
\end{align}
We stabilize this $\Omega$  
via the stationary condition set for given $\theta$, $\mu_B$, $\mu_5$, and $T$:  
$ 
\frac{\partial\Omega}{\partial S^{\prime}}=\frac{\partial\Omega}{\partial P^{\prime}}=0
$.

As is evident in the form of Eq.(\ref{eq:9}), 
$\mu_5$ and $\theta$ 
%do not essentially separate two thermal correction terms, 
%the second and third terms in square bracket, because do not break the $C$ symmetry. 
%Therefore, those two sources are not significant in addressing 
%the chiral phase transition features in a view of the scale violation classes: 
%they 
merely act as a catalyzer against the CW-type scale anomaly, i.e., the driving force for the phase transition in the case without them. 
For instance, even for a large $\mu_5$ at $T=\mu_B=0$, 
in the case with massive quarks including physical point, $\mu_5$ does not generate the Debye mass or logarithmic term for $M$, hence 
$\mu_5$ itself does not trigger the chiral restoration~\footnote{ 
In the chiral limit, $\mu_5$ can act like $\mu_B$ as seen from Eq.(\ref{Es-M}), 
and therefore can trigger the first order phase transition as has been discussed in the literature~\cite{Chernodub:2011fr}.  
Even at physical point, a dynamically induced $\mu_5$ as the mean field of the axialvector field can also serve as 
the source for triggering the first order phase transition~\cite{Shao:2022oqw}, which we will not consider in the present study. 
%In the present work, we will not consider this possibility just for simplicity.
}. 
The NJL-MFA with nonzero $\mu_5$, however, involves the high regularization scheme dependence~\cite{Yu:2015hym}. 
In the literature~\cite{Farias:2016let} a proper subtraction scheme has been shown to  support the chiral crossover, not the first order, even in the chiral limit, where $\mu_5$ acts as a catalyzer to enhance 
the chiral/axial breaking via efficiently driving the topological charge fluctuation to the axial anomaly.  
We will come back to this point later on.

\section{Exploring CW-type chiral phase transition} 
\label{sec3-main}

In this section, we discuss the features of the chiral phase transition based on the thermodynamic potential in Eq.(\ref{eq:9}) in light of realization of the CW-type phase transition on the basis of the scale violation classification. 
The cases we shall study are classified into two: with $\mu_B=0$ (Sec.~\ref{sec3}) or $\mu_B \neq 0$ (Sec.~\ref{sec4}). 
Our main phenomenological proposal is in the former case, where an ALP is predicted to deform the conventional chiral crossover/phase transition into the CW-type supercooling, which we call the ALP-assisted CW-type chiral phase transition.  
This particular scenario will separately be addressed in more detail in the later 
section, Sec.~\ref{ALP}.

\subsection{The case with $\mu_B=0$} 
\label{sec3}

In this case Eq.(\ref{eq:9}) goes like 
\begin{equation}\label{eq:11}
    %\Omega\Bigg|_{\mu_B=\mu_5=0}
    \left.\Omega\right|_{\mu_B=\mu_5=0}=2g_{s}(S^{\prime 2}+P^{\prime 2})+2g_{d}(S^{\prime 2}-P^{\prime 2})-2 N_{c}N_{f}\int\frac{d^{3}p}{(2\pi)^{3}}\left[E_{s=0}+2T\text{ln}(1+e^{-\frac{E_{s=0}}{T}})\right]
\,. 
\end{equation}
It is well known in the two-flavor NJL model that at the physical point for quark masses, the chiral phase transition at vanishing baryon chemical potentials is crossover (see, e.g., a review~\cite{Hatsuda:1994pi}). 
This is also true even with nonzero $\theta$, as long as $\theta \neq \pi$, in which case the CP phase transition along $P=S'$ undergoes like a second order~\cite{Boer:2008ct,Boomsma:2009eh,Sakai:2011gs,Sasaki:2013ewa,Lu:2018ukl,Huang:2024nbd}. 
Now we understand this phase transition property in terms of the scale violation classes. 

First, we take $\theta=0$ for simplicity. 
It turns out that this will not essentially lose generality on the phase transition property 
even with $\theta \neq 0$, as long as $0<  \theta < \pi$.  
With $\theta = 0$, we have $S' = S$ and $P'=P=0$, or equivalently, 
$\beta=0$ in Eq.(\ref{eq:8}), hence $M^2 = \alpha^2$ in Eq.(\ref{Es-M}). 
In that case Eq.(\ref{eq:11}) can be simplified further to 
\begin{equation}\label{theta0}
    %\Omega\Bigg|_{\mu_B=\mu_5=0}^{\theta=0}
    \left.\Omega\right|_{\mu_B=\mu_5=0}^{\theta=0}=\frac{(M-m)^2}{2 (g_{s} + g_d)}-2 N_{c}N_{f}\int\frac{d^{3}p}{(2\pi)^{3}}\left[E_{s=0}+2T\text{ln}(1+e^{-\frac{E_{s=0}}{T}})\right]
\,. 
\end{equation}
Now, this is simply a function of $M$.

Consider first the vacuum case with $T=0$. 
The tree-level (or bare) mass term, $(M - m)^2$, has come from the four-fermion interactions 
along with $g_s$ and $g_d$, 
including the shift by the current quark mass. 
This is a soft-scale breaking term. 
In addition, the first term in the loop correction part (inside the square brackets in Eq.(\ref{theta0})) includes quadratic $\sim M^2 \Lambda^2$ and logarithmic divergent terms $\sim M^4 \ln(\Lambda/M)$, where $\Lambda$ denotes the loop momentum cutoff. 
The former definitely takes a negative sign and gives a destructive contribution to the bare $M^2$ term, which destabilizes the chiral symmetric phase at $M=0$ up to 
the linear shift by $m$, hence triggers the dynamical chiral symmetry breaking. 
This is another soft-scale breaking term. 
The net soft-breaking term takes a largely negative sign at $T=0$.  
Meanwhile, the latter logarithmic divergent term takes nothing but a form 
of the CW-type scale anomaly. 
Thus, at $T=0$, the effective potential derived from Eq.(\ref{theta0}) (with renormalization prescribed) can be cast 
into essentially the same form as the generic CW-type potential given in Eq.(\ref{essential}), 
with a large negative curvature around $M=0$ ($m_0^2 <0$ in Eq.(\ref{essential})): 
\begin{align}
     %\Omega\Bigg|_{\mu_B=\mu_5=0}^{\theta=0, T=0} 
     \left.\Omega\right|_{\mu_B=\mu_5=0}^{\theta=0, T=0}\approx \frac{(M-m)^2}{2 (g_{s} + g_d)}-\frac{N_f N_{c}}{4 \pi^2} 
     \left( M^2 \Lambda^2 - \frac{M^4}{4} \ln\frac{\Lambda^2}{M^2} \right)  
\,, \label{T0}
\end{align}
for $\Lambda \gg M$.

At finite $T$, the negative curvature around $M=0$ gets thermal corrections 
(the second term in the square brackets in Eq.(\ref{theta0})). 
This starts to act as a cancellation of the soft-scale breaking term $\propto M^2$. 
At some high enough $T$, the original soft-breaking term, i.e. the dynamical chiral symmetry breaking will be canceled (i.e., $m_0^2=0$ in Eq.(\ref{essential})), 
which determines the typical critical temperature $T_c$.  
As $T$ gets further higher, the soft-breaking term is instead developed by 
the thermal mass term $\sim T^2 M^2$. 
At this moment, one might suspect that a potential barrier is built around 
$M=0$ by the positive curvature (the thermal mass) $\sim  (+ T^2 M^2) $.  
%and the logarithmic damping term 
%$\sim  (- M^4 \ln M)$, just like the case of the CW-type potential at high $T$, 
%hence the thermal chiral phase transition goes like first order or more like 
%a supercooling form. 
%However, it is not the case because of 
In addition, 
the original CW-type logarithmic term is actually canceled by
the high $T$ correction. 
To see this cancellation more explicitly, we work on the high $T$ expansion for 
the thermal correction term in Eq.(\ref{theta0}), which goes like  
\begin{align} 
%\label{eq:14}
\left.-2 N_c N_f  \cdot 2 T \int\frac{d^{3}p}{(2\pi)^{3}} \ln(1+e^{-\frac{E_0}{T}})  
\right|_{T\gg M}
&= 
N_c N_f \left[-\frac{7\pi^{2}}{180}T^{4}+\frac{1}{12}T^{2}M^{2}+\frac{1}{16\pi^{2}}M^{4}\ln M^2 -\frac{1}{16 \pi^{2}}M^{4}\ln a_{f}T^{2}
+ \cdots 
\right] 
\,, 
\label{high-T-exp}
\end{align}
where $a_f = 2 \ln \pi - 2 \gamma_E \simeq 1.14$. 
As seen from Eq.(\ref{T0}), 
the original CW-type logarithmic term $\sim M^4 \ln M$ is precisely canceled.

%Thereby, the CW-type first-order phase transition 
%is not realized: no barrier is generated. 
%This is completely consistent with the well-known fact that 
%the two-flavor chiral phase transition in the MFA-NJL model, 
%with $\mu_B=0$, is second order in the chiral limit with 
%$m=0$, and crossover at the physical point. 
%This cancellation phenomenon can also be understood in view of 
%the original scale anomaly cancellation in an extremely hot medium, 
%which implies a free quark gas picture simply following 
%the Stephan-Boltzmann form of pressure, $\Omega \sim p \sim T^4$ (corresponding to 
%the first term in Eq.(\ref{high-T-exp})). 
%This cancellation is further promoted to work with nonzero $\mu_5$, because 
%$\mu_5$ can act as a catalyzer for the chiral/axial breaking when 
%a proper regularization scheme is carried out~\cite{Farias:2016let}. 

This scale anomaly cancellation does not happen 
when the soft-scale breaking mass is small enough as in the case that 
has been addressed in the literature~\cite{Gupta:2011ez}. 
This is because the high $T$ expansion will not accurately approximate the thermal potential 
due to the scale invariance around $M=0$. 
%hence a barrier is created and 
%the phase transition undergoes like first order and supercooling. 
Thus, the scale anomaly cancellation is intrinsic to a negatively 
large soft-scale breaking case, which makes the curvature of the potential 
deeply negative around $M=0$, 
like in the present NJL scenario with significant dynamical 
chiral symmetry breaking.

Even when nonzero $\theta$ is turned on, 
we can again arrange $\Omega$ so that  
the thermal loop and vacuum-one loop terms are still singly 
dependent of $M$. 
Nonzero $\theta$ effects only come into the tree-level part, 
in which either $P'= \frac{m \sin \frac{\theta}{2} - \beta}{2(g_s - g_d)}$ or $S'=\frac{m \cos \frac{\theta}{2} - \alpha}{2(g_s + g_d)}$ direction can be chosen as the actual mean field directions in addition to the $M = \sqrt{\alpha^2 + \beta^2}$ direction. Thus, dynamical chiral symmetry breaking is still monitored along 
the $M$ direction, keeping the size of the soft-scale symmetry breaking at each $T$. 
Hence the high $T$ expansion works as in the case with $\theta=0$ and the scale anomaly cancellation is still operative, so the  
chiral phase transition substantially follows 
the same way. 
%and no first order phase transition can be seen even with $\theta\neq 0$. 
On the other hand, the $CP$ phase transition feature 
observed along $S'$ or $\alpha$ highly depends on $\theta$, which exhibits 
a second order type when $\theta=\pi$~\cite{Boer:2008ct,Boomsma:2009eh,Sakai:2011gs,Huang:2024nbd}.

The dynamical chiral symmetry breaking occurs when the gap equation, i.e., 
the stationary condition for Eq.(\ref{T0}) gets nontrivial solution. 
It is the case where the following condition is met:  
\begin{align}
    \frac{N_f N_c (g_s + g_d) \Lambda^2}{2 \pi^2} \ge  1 
\,,   \label{ratio}
\end{align}
in the chiral limit, for simplicity. 
At the criticality where $\frac{N_f N_c (g_s + g_d) \Lambda^2}{2 \pi^2} =1$, 
the soft-scale breaking mass becomes zero: $m_0^2=0$ in terms of Eq.(\ref{essential}).  
In the two-flavor QCD case, we have $\frac{N_f N_c (g_s + g_d) \Lambda^2}{2 \pi^2} \sim 1.5$, 
which is fixed by fitting to QCD hadron observables  
(see also Eq.(\ref{para})). Thus, QCD indeed yields a significantly large soft-scale breaking as dynamical chiral symmetry breaking. 
Note that the near criticality condition, $\frac{N_f N_c (g_s + g_d) \Lambda^2}{2 \pi^2} \sim 1$, also implies the dynamics near an ultraviolet fixed point (in the chiral broken phase for $\frac{N_f N_c (g_s + g_d) \Lambda^2}{2 \pi^2}>1$),  
$\Lambda \frac{\partial [(g_s + g_d)\Lambda^2]}{\partial \Lambda} \sim 0$, 
i.e., the almost quantum scale invariance.

In the MFA, the near criticality condition for soft-scale breaking, 
$\frac{N_f N_c (g_s + g_d) \Lambda^2}{2 \pi^2} =1$, 
implies 
\begin{align}
    \frac{M \Lambda^2}{\langle - \bar{q}q  \rangle/N_f} = \frac{4 \pi^2}{N_c} \sim 12  \times \left( \frac{3}{N_c} \right)
    \,, \label{critical}
\end{align}
where we have used Eq.(\ref{eq:8}) with $\theta=m=0$. 
Two-flavor QCD gives $\frac{M \Lambda^2}{\langle - \bar{q}q  \rangle/N_f} \sim 20$. 
For a fixed place of $S = \langle \bar{q}q  \rangle$ in the thermodynamic potential, 
Eq.(\ref{critical}) requires a smaller dynamical (full) mass $M$ by about 40\%, compared to the existing two-flavor QCD case. 
Since generically $\langle - \bar{q}q  \rangle \sim M^3$ (when evaluated at the renormalization scale $\mu=M$), the condition in Eq.(\ref{critical}) may be read as 
$\Lambda/M \gg 1$. 
This indicates a large scale hierarchy between the ultraviolet and infrared scales 
intrinsic to the underlying theory contrast to QCD: ordinary QCD yields almost the same order for them, $\Lambda \sim (2 -3) M$. 
Soft-scale breaking is ensured by the nature of the 
almost scale-invariant gauge theory, and the CW-type scale anomaly is induced 
by nonperturbative running of the gauge coupling associated with dynamical chiral symmetry breaking~\cite{Leung:1985sn}. 
One candidate theory to realize such a large scale gap is many flavor QCD, 
e.g., QCD with $N_f=8$~\cite{Aoki:2013xza,LSD:2014nmn,Hasenfratz:2014rna}. 
In this scenario, the approximate quantum scale symmetry gets anomalous dominantly by the CW-type scale anomaly of the form $\sim M^4 \ln M$. 
This characteristic feature has so far been applied to various phenomena, 
such as dynamical CW inflation scenarios~\cite{Ishida:2019wkd,Cacciapaglia:2023kat,Zhang:2023acu,Zhang:2024vpp,Liu:2024xrh,Zhang:2025kbu,Cacciapaglia:2025xqd}, 
models of dynamical electroweak symmetry breaking~\cite{Matsuzaki:2012gd,Matsuzaki:2012vc,Matsuzaki:2012mk,Matsuzaki:2013eva,Matsuzaki:2015sya,Kasai:2016ifi,Hansen:2016fri,Appelquist:2017wcg,Appelquist:2017vyy,Cata:2019edh,Appelquist:2019lgk,Brown:2019ipr,Elander:2025fpk,Faedo:2024zib,Appelquist:2024koa,Ingoldby:2023mtf,Freeman:2023ket,LSD:2023uzj,Choi:2011fy,Choi:2012kx,Appelquist:2022mjb,Appelquist:2022qgl,Golterman:2020utm,Miura:2018dsy}, 
and phenomenological applications to QCD dilaton~\cite{Kawaguchi:2020kce,Crewther:2013vea,Li:2016uzn,Fujii:2025aip,Sheng:2025jvn,Zhang:2024iye,Zhang:2024sju,Bersini:2023uat,Ma:2023ugl,Shao:2022njr,Cata:2019edh,Li:2017udr,Paeng:2017qvp,Ma:2016gdd,Ma:2016nki}.

In the following, we will propose a new scenario for $\mu_B \sim  0$, 
which realizes the approximate scale invariance only at around $T=T_c$ (Sec.~\ref{ALP}). 
Before this proposal, in the next section 
we shall take a look at the case with nonzero $\mu_B$, where the realization of the supercooling chiral phase transition, as has been discussed 
in the literature~\cite{Shao:2024dxt,Shao:2024ygm}, will be recapped in terms of 
the class of the soft-scale breaking and the Debye mass effect.

\subsection{The case with $\mu_B\neq 0 $}
\label{sec4}

At an extreme limit where $T \sim 0$ and finite $\mu_B$ (with $\mu_5=0$),  
%Under the conditions of zero temperature and vanishing chiral chemical potential, 
the finite-density correction to $\Omega$ in Eq.(\ref{eq:9}) can be evaluated as 
\begin{align} 
%\Omega_{\mu_{B}-1loop}&=
& -2 N_c N_f \lim_{T\to0}T\int\frac{d^{3}p}{(2\pi)^{3}}\left[\text{ln}\left(1+e^{-\frac{E_0+\frac{1}{3}\mu_{B}}{T}}\right)+\text{ln}\left(1+e^{-\frac{E_0-\frac{1}{3}\mu_{B}}{T}}\right)\right] 
 \notag\\ 
%\nonumber\\
%&=-\lim_{T\to0}T\int\frac{d^{3}p}{(2\pi)^{3}}\text{ln}\left(1+e^{-\frac{E-\frac{1}{3}\mu_{B}}{T}}\right)\nonumber\\
%&=\int_{0}^{p_{E}=\sqrt{(\frac{1}{3}\mu_{B})^{2}-M^{2}}}\frac{p^{2}}{2\pi^{2}}\left(E-\frac{1}{3}\mu_{B}\right)dp\nonumber\\
&\approx \frac{N_cN_f}{24\pi^{2}}\left(- (\frac{1}{3}\mu_{B}) \sqrt{(\frac{1}{3}\mu_{B})^{2}-M^{2}}\left(2\cdot(\frac{1}{3}\mu_{B})^{2}-5M^{2}\right)\right)\nonumber\\
& + \frac{N_c N_f}{24\pi^{2}}\left(- 3M^{4}\ln{\frac{(\frac{1}{3}\mu_{B})+\sqrt{(\frac{1}{3}\mu_{B})^{2}-M^{2}}}{M}}\right)\nonumber\\
&\equiv \Omega_{\mu_{B}}^{(1)}+\Omega_{\mu_{B}}^{(2)}
\,. 
\end{align}
The first term defined as $\Omega_{\mu_B}^{(1)}$, up to a term constant in $M$, 
goes like 
$\sim \frac{N_cN_f}{12 \pi^2} (\frac{\mu_B}{3})^2 M^2$ for a large $\mu_B\gg M$, which is the Debye mass screening. 
The second term 
%, $\Omega_{\mu_B}^2$, 
%\begin{align} 
$
\Omega_{\mu_B}^{(2)} 
\sim \frac{-N_c N_f}{16 \pi^2} M^4 \ln \frac{ \left[(\frac{\mu_B}{3})+ \sqrt{(\frac{\mu_B}{3})^2-M^2} \right]^2}{M^2}
$
%\,, 
%\end{align} 
takes a scale anomaly form of the CW type and, in an extreme limit $\mu_B \gg M$, 
precisely cancels the original one at $\mu_B=0$ in Eq.(\ref{theta0}).  
A potential barrier is generated at a moderately large $\mu_B$ by 
the creation of the Fermi surface, acting as a repulsive force, 
dictated by $\Omega_{\mu_B}^{(1)}$. 
This triggers the first order phase transition at $T=0$.

At finite $T$, the cancellation for the original scale anomaly form of the CW-type 
becomes incomplete, hence it will survive during the thermal phase transition, as long as 
the Debye mass screening by $\mu_B$ efficiently works so as to make the dynamical 
chiral symmetry breaking milder. 
Thus, a large enough $\mu_B$ can control the size of the soft-scale breaking. 
Still at finite $T$, thus the first order phase transition can be seen 
because of the residual CW-type scale anomaly effect.

\begin{figure}[t]
    \centering
    %\begin{subfigure}[t]{0.48\linewidth}
    %    \centering
    %    \includegraphics[width=\linewidth]{figures/notefigures/potential-high-T.pdf}
    %    \caption{High $T$}
    %\end{subfigure}
    %\hfill
    %\begin{subfigure}[t]{0.48\linewidth}
    %    \centering
        \includegraphics[width=0.6\linewidth]{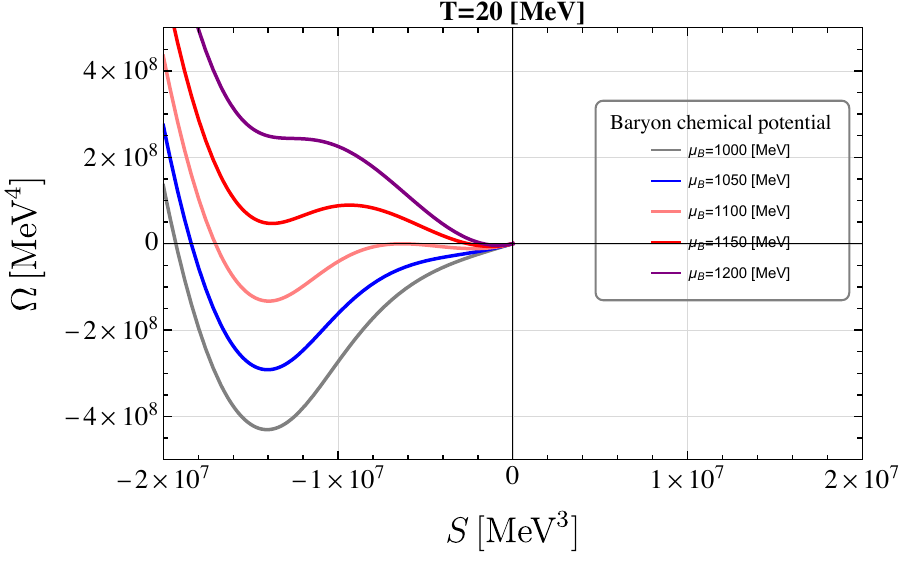}
    %    \caption{Low $T$}
    %\end{subfigure}
    \caption{The variation of the thermodynamic potential with respect to $\mu_B$ at low $T$ ($T=20$ MeV), as a function of $S = \frac{M - m}{2(g_s + g_d)}$ at $\theta=0$, 
    in the NJL-MF description monitoring two-flavor QCD at physical point (for details, see the text). } 
    \label{fig:2}
\end{figure}

\begin{figure}[t] 
\includegraphics[width=0.6\linewidth]{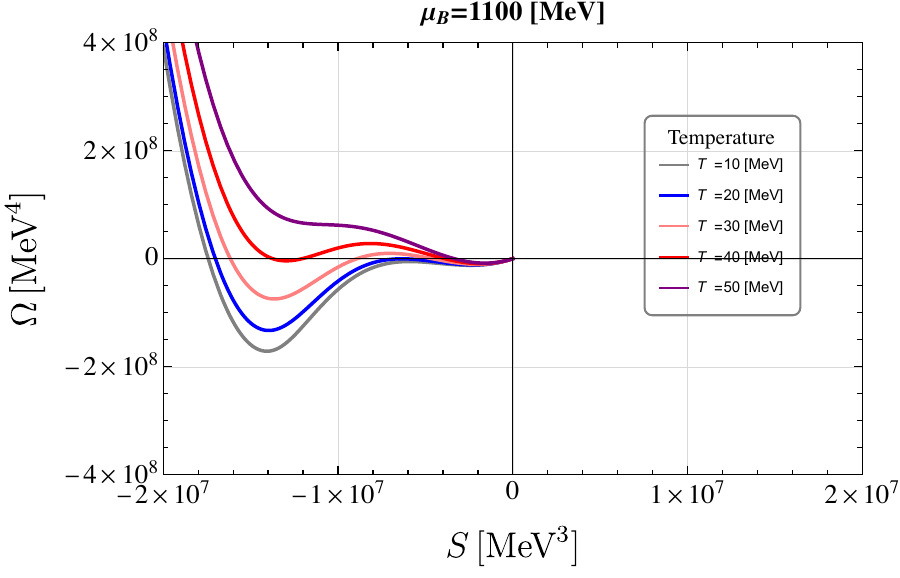}
\caption{The thermal chiral phase transition of the CW type including 
a soft-scale breaking mass term induced from the Debye mass at $\mu_B=1100$ MeV and dynamical chiral symmetry breaking. The model parameters have been setup at physical point as in Fig.~\ref{fig:2}. } 
\label{muB1100MeV}
\end{figure}

Figure~\ref{fig:2} demonstrates how the CW-type potential can be realized by finite $\mu_B$ at low enough $T$. 
The desired size of $\mu_B$ would thus be $\sim 1100$ MeV~\footnote{
 When $\mu_B \sim 1100$ MeV, we have the quark chemical potential 
$\mu_q = \mu_B/3 \sim 360$ MeV. One might think that this $\mu_q$, which is close to the cutoff 
$\Lambda \simeq 590$ MeV (as in Eq.(\ref{para})), would lead to 
poor accuracy and a cutoff artifact in the results on the chiral phase transition. 
However, it is not the case: 
the dynamical quark mass $M =2(g_s+g+d)|S| \sim 364$ MeV, which  is read off from Eq.(\ref{para}) and the nontrivial vacuum place in Fig.~\ref{muB1100MeV}. From these values, 
we estimate the Fermi momentum $p_F = \sqrt{\mu_q^2 - M^2}$ to be $\sim 47$ MeV. This is the 
scale that plays the role of the effective cutoff for the medium effect at $T\sim 0$, above which the higher momentum contributions are exponentially suppressed. 
At finite $T$, the suppression is still exponential: for momenta $p\gg \mu_q, T$, 
the integrand for the medium part of $\Omega$ in Eq.(\ref{eq:9}), $\Omega(T, \mu)-\Omega(T,0)$, 
scales as $e^{- (p-\mu_q)/T}$, making contributions from $p \sim \Lambda$ almost negligible for $T$ of our interest ($\sim 100$ MeV). Thus, the medium correction to the chiral phase transition 
is utterly free and decoupled from the cutoff dependence, as long as the Fermi momentum $<\Lambda$, as in the present case. 
}.
Requiring this size of $\mu_B$ is also consistent with the analysis in the literature~~\cite{Shao:2024dxt,Shao:2024ygm}. 
Note that in Fig.~\ref{fig:2}, 
the soft-scale breaking effect $\propto S^2$ contributes negatively around $S \sim 0$. 
Here, to fix the model parameters, we have referred to empirical hadron observables in the isospin symmetric limit 
at vacuum~\cite{Boer:2008ct,Boomsma:2009eh,Sakai:2011gs}: the pion mass $m_\pi=140.2\, \mathrm{MeV}$, the pion decay constant $f_\pi=92.6\, \mathrm{MeV}$, and quark condensate $\langle\bar{u} u\rangle=\langle\bar{d} d\rangle=(-241.5 \,  \mathrm{MeV})^3$. 
Then the model parameters can be determined as~\cite{Boer:2008ct,Boomsma:2009eh,Sakai:2011gs}, 
\begin{align}
& m\simeq 6 \, \mathrm{MeV},  \qquad \Lambda \simeq 590 \, \mathrm{MeV} \,, \notag \\ 
&g_s=2(1-c) G_0, \qquad g_d=2 c G_0 \,, \qquad {\rm with} \qquad 
c\simeq 0.2\,, \qquad G_0 \Lambda^2 \simeq 2.435
\,. \label{para}
\end{align}

In Fig.~\ref{muB1100MeV} we show the thermal evolution of the CW-type chiral phase transition with $\mu_B =1100$ MeV fixed. 
The CW-type supercooling can indeed be achieved in this setup. 
The stochastic GW background can then be produced and the predicted spectra 
could be similar to those discussed in~\cite{Shao:2024dxt,Shao:2024ygm}, where 
a two-flavor quark meson model with a large $\mu_B$ such as $\mu_B \sim 1$ GeV has been shown to generate nano hertz GW signals, without clarifying the limit to 
the CW-type chiral phase transition. 
As inspired also in the literature~\cite{Shao:2024dxt,Shao:2024ygm}, 
the Affleck-Dine baryogenesis~\cite{Affleck:1984fy,Linde:1985gh,Dine:2003ax} could produce such a large $\mu_B$ in the QCD phase transition epoch.

\section{ALP assisted CW-type chiral phase transition in QCD} \label{ALP}

In this section, we propose a new scenario to realize 
the CW-type supercooling-chiral phase transition in the QCD thermal history. 
This scenario is categorized into the case with $\mu_B=0$.

\subsection{Cancellation of soft-scale breaking mass term in QCD plasma} 

We start with considering a new scalar field ($\Phi$) coupled to $M$, like $\Phi^2 M^2$ of the portal form, and allow $\Phi$ 
to develop the vacuum expectation value and generate an extra soft-scale breaking mass term, to cancel the dynamical chiral symmetry breaking at low-enough $T$.  
Suppose that this vacuum expectation value becomes nonzero in the chiral symmetric phase 
at high $T$, and almost vanishes at vacuum with $T=0$. 
This setup ensures that the extra soft-scale breaking effect is activated to trigger the cosmological CW-type phase transition at $T=T_c$, and below $T_n$, it freezes out 
to make the QCD-chiral broken phase normal, so that the present-day QCD hadron physics is intact.

The decoupling of extra soft-scale breaking can be realized, say, 
if the vacuum expectation value of $\Phi$ scales with $\mu_5$, the chiral chemical 
potential, so the $\Phi$ field acts like an ALP coupled to the axial charge, 
as has been discussed in~\cite{Zhang:2025kbu}. 
This decoupling is then encoded with the freezing out of the QCD sphaleron, which would happen in the epoch of the confinement-deconfinement phase transition, 
which takes place seemingly simultaneously with the chiral phase transition. 
As long as the QCD sphaleron is active in hot QCD plasma, the chiral chemical potential $\mu_5$ along with a local CP-odd domain is created~\cite{Manton:1983nd,Klinkhamer:1984di}, so that the QCD vacuum, characterized by 
the strong CP phase $\theta$, and the vacuum fluctuation (in the spatial-homogeneous direction) become significantly sizable~\cite{Kharzeev:2007tn,Kharzeev:2007jp,Fukushima:2008xe} within the QCD time scale~\cite{McLerran:1990de,Moore:1997im,Moore:1999fs,Bodeker:1999gx}: 
the sphaleron transition rate is not suppressed by the thermal effect in contrast to 
the instanton's one~\cite{Moore:1997im,Moore:1999fs,Bodeker:1999gx}. 
The time fluctuation $\partial_t \theta(t)$, to be referred to as the chiral chemical potential, $\mu_5$~\cite{Kharzeev:2007tn,Kharzeev:2007jp,Fukushima:2008xe,Andrianov:2012hq,Andrianov:2012dj}, will be significant as well when the non-conservation law of the $U(1)$ axial symmetry is addressed.

This $\mu_5$ arises as the static solution for the quasi thermal equilibrium system~\cite{McLerran:1990de}, and its thermal evolution around the QCD phase transition involves a nonperturbative decoupling of the QCD sphaleron due to the confinement. which makes $\mu_5$ frozen out as well.    
%In the present work, we shall leave the issue on the nonperturbative $T$-dependence of $\mu_5$ to another publication, and 
Therefore, we will take $\mu_5$ to be constant in $T$ to be of ${\cal O}(100)$ MeV, as has conventionally been applied in the $\mu_5$ physics in the chiral phase transition~\footnote{\label{foot}
The constant $\mu_5$ can be understood as follows:  
The chiral chemical potential $\mu_5$ in terms of thermodynamics is related to 
the axial number density $n_5$ and the axial susceptibility $\chi_5$ as $\chi_5 = \frac{\partial n_5}{\partial \mu_5}$. 
In the thermal equilibrium, the nonconservation law of $n_5$ yields 
$n_5 \sim \Gamma_{\rm sph}/\Gamma_{\rm chi}$, where 
$\Gamma_{\rm sph}$ is the QCD sphaleron-transition rate and $\Gamma_{\rm chi}$ denotes the chirality-flip rate $\sim  N_c^2 T^3/M^2(T)$ with the dynamical quark mass $M(T)$, which, e.g., arises via 
the gluon - quark elastic scattering, $q + g \leftrightarrow q + g$. 
%At high enough $T (\gg M(T))$, $\Gamma_{\rm chi} = n_g \cdot \langle  \sigma v \rangle_T$ can %roughly be $\sim T^3 \cdot \alpha_s \frac{M^2(T)}{T^4}$, where the first factor ($T^3$) comes %from the thermal number density of gluons ($n_g$), while the second factor ($\alpha_s M^2(T)/%T^4$) from the thermal averaged cross section $(\langle \sigma v  \rangle_{T})$. 
%Therefore, $n_5$ scales with $T$ as $n_5 \sim T^5/M^2(T)$. 
Therefore, the chiral chemical potential thermally evolves like $\frac{d \mu_5}{dT} = \frac{dn_5}{dT}\frac{1}{\chi_5} \sim \frac{d}{dT} (\frac{T\cdot \Gamma_{\rm sph}}{M^2(T)}) \frac{1}{\chi_5}$. 
Around the chiral criticality ($T \sim T_c$), $(1/M(T))$ peaks at $T\simeq T_c$. 
Since the $U(1)$ axial and $SU(2)$ chiral restorations are contaminated, and $n_5$ acts as the indicator for the $U(1)$ axial breaking strength -- i.e,. $n_5 \to \infty$ when the $U(1)$ axial symmetry is restored --, $n_5$ also exhibits a peak structure around $T \sim T_c$. 
This is a pseudocriticality when the theory is at the physical point, which follows a power law reflecting the remnant of the universality class (e.g. $O(4)$) around $T\sim T_c$. 
The QCD sphaleron rate $\Gamma_{\rm sph}$ exhibits a Boltzmann suppression 
$\sim \frac{N_c^2-1}{N_c^2}\cdot (\alpha_s N_c)^5 \cdot  T^4 e^{- \Lambda_{\rm QCD}/T}$~\cite{McLerran:1990de,Moore:2010jd} due to the mass gap generation (confinement). Thus, $\frac{d \mu_5}{dT}  \sim 0$ around $T\sim T_c$ (for $T>T_c$). When the theory undergoes (the CW-type) ultra supercooling, as in the present model, below $T<T_c$, $\chi_5$ and $M$ will not drastically change with $T$ to be almost saturated respectively to the vacuum values $\sim m_{\eta'}^2$ and $\sim (\langle  - \bar{q}{q} \rangle_{T=0})^{1/3}$, where $m_{\eta'}$ is the QCD $\eta'$ mass associated with the $U(1)$ axial anomaly at the vacuum. 
On the other hand, $\Gamma_{\rm sph}$ still keeps Boltzmann damped when $T$ cools down til the nucleation/percolation temperature $T_n\ll T_c$. Therefore, we can conclude that
$\frac{d \mu_5}{dT}  \sim 0$ even in the case of the supercooled chiral phase transition. }.

We assume an axial scalar field $\Phi = \phi_s + i \phi_p$, which we call the $\Phi$-ALP field, to couple to up and down quarks via the Yukawa coupling term, 
\begin{align}
    y  \, (\bar{q}_L q_R \Phi + \bar{q}_R q_L \Phi^*) = y (\bar{q}q \phi_s + \bar{q}i \gamma_5 q \phi_p)  
\,,   \label{yS}
\end{align}
with $y \ll 1$. 
Then, the $\Phi$ coupling to the chiral chemical potential $\mu_5$  
would also be induced through the $U(1)$ axial covariance: 
\begin{align}
    |D_\mu \Phi|^2 \,, 
    \qquad {\rm with} \qquad 
    D_\mu \Phi = (\partial_\mu  +  2 i \mu_5 \delta^0_\mu) \Phi 
    \,. 
\end{align}
Hence $\Phi$ gets the effective potential along with $\mu_5$ 
\begin{align} 
  V_{\mu_5}(S) = - (2 \mu_5)^2 |\Phi|^2
\,. \label{mu5-pot}
\end{align}
This is definitely a negative mass squared term ensured by the gauge covariance for the external axialvector field $A_\mu = \mu_5 \delta_\mu^0$, which corresponds to 
the repulsive force from the vector interaction against the scalar probes charged under the corresponding gauge.

Thus, combined with the quartic term $\lambda |\Phi|^4$, $\Phi$ can get 
nonzero vacuum expectation value, $v_\Phi = \frac{2 \mu_5}{\sqrt{\lambda}}$. 
We assume that the $\Phi$ field settles at the vacuum expectation value at a high enough $T$, even when 
$T \gtrsim T_c$, where the QCD sphaleron actively generates $\mu_5$. 
At $T \sim T_c$, the Yukawa coupling term in Eq.(\ref{yS}), when applied to the current NJL-MF  framework, 
modifies the thermodynamic potential to deform the type of the chiral phase transition. 
It makes a shift of the mean field terms of the scalar ($S$) and pseudoscalar ($P$) only in the loop correction parts, by $(y \, \phi_s)$ and $(y \, \phi_p)$, respectively. 
Without loss of generality, we take $\langle \phi_s \rangle = v_\Phi$ and $\langle \phi_p \rangle =0$. 
Keeping $\theta=0$, hence $P=0$, we then find that up to the thermal correction parts, 
the vacuum contribution in Eq.(\ref{T0}) looks like modified as  
\begin{align}
    \Omega 
    &\approx 
    \frac{{\cal S}^2}{2(g_s + g_d)} 
   % + \frac{g_d P^2}{g_s^2-g_d^2} 
   %\notag\\ 
   %&
   - \frac{N_fN_c}{4\pi^2} 
   \left[
   \widetilde{S}^2 \Lambda^2 
   - \frac{(\widetilde{S}^2)^2}{4} 
   \ln \frac{\Lambda^2}{\widetilde{S}^2}  \right]
   \,, \label{Phi-shift}
\end{align}
where 
\begin{align}
\widetilde{S}^2
&\equiv 
(y \, v_\Phi - {\cal S})^2 
\,, \notag \\ 
{\cal S}^2 & 
\equiv 
4 (g_s + g_d)^2 S^2 \,, 
\label{Scal}
\end{align}
In Eq.(\ref{Phi-shift}) we have taken the chiral limit $m\to 0$ for simplicity. 
Since the Yukawa coupling in Eq.(\ref{yS}) is CP invariant, 
we assume $\langle P \rangle =0$ at $\theta=0$, as in ordinary QCD. 
We find the extra positive mass-squared contribution to ${\cal S}$,  
\begin{align}
   %m_0^2|_{\widetilde{P}} 
   \sim   \frac{3 N_f N_c}{8 \pi^2} (y v_\Phi)^2 %\frac{y^2 \mu_5^2}{\lambda}  
   = \frac{3 N_f N_c}{2 \pi^2} \frac{y^2 \mu_5^2}{\lambda} 
\,, 
\end{align}
up to the logarithmic factor of ${\cal O}(1)$. 
Thus, the total ${\cal S}^2$ term goes like 
\begin{align}
  \Omega_{{\cal S}^2} 
  &\sim \left[ \frac{1}{2(g_s+g_d)} - \frac{N_fN_c}{4\pi^2} \Lambda^2 + 
    \frac{3 N_f N_c}{8 \pi^2} (y v_\Phi)^2  \right] {\cal S}^2 
    + \cdots 
\,, \notag\\ 
& \sim \frac{3 N_f N_c}{8 \pi^2} \left[  (y v_\Phi)^2 - 
 (274\,{\rm MeV})^2    \right] {\cal S}^2 + \cdots 
\,, \label{shift-by-mu5}
\end{align}
where ellipses represent thermal correction terms at $T\gtrsim T_c$, and we have quoted values of the model parameters referenced in Eq.(\ref{para}). 
As clearly seen from Eq.(\ref{shift-by-mu5}), 
the soft-scale breaking mass term can be canceled when $(y v_\Phi) \sim 274 $ MeV.

The cancellation in Eq.(\ref{shift-by-mu5}) is thus dedicate 
and actually the resultant CW-type scale anomaly takes a wrong sign 
for generation of the first order phase transition at this point. 
To make it more complete, we may incorporate the meson fluctuation contribution, 
beyond the MFA level. 
Although the meson loop corrections should generically be subleading and decoupled for ${\cal S}/\Lambda \sim 1$ because the meson mass is typically given as $\sim \sqrt{\lambda_{\rm meson}} {\cal S} \ll \Lambda$ with $\lambda_{\rm meson} \gg 1$, 
they can play a crucial role around the criticality (at $T\sim T_c$). In particular, when the negative mass square of the chiral order parameter ${\cal S}$ is turned off as in the present scenario, the pion-loop can give a significant CW-type scale anomaly of the form $\sim + {\cal S}^4 \ln {\cal S}$ with the right sign.  This term can be a trigger for dynamical chiral symmetry breaking to determine the curvature and barrier structure around the true vacuum where the net $m_\pi \sim 0$, 
even away from the origin of the potential (${\cal S}\sim 0$)~\footnote{ 
This feature is characteristic of the present scenario, in contrast to 
the conventional one with a large negative mass squared at the ``tree-level" of the meson sector, which dominates because all the field-dependent meson masses, including the pion mass, become heavy enough due to the large mass square, and the loop corrections are highly suppressed, when the curvature and/or barrier structure of the potential around the true vacuum is determined.}.  
In what follows, we will more explicitly see this pion-loop trigger.

The present NJL-MFA dynamically generates the meson sector Lagrangian, which takes the 
form 
\begin{align}
    {\cal L}_{\rm meson} 
    = \frac{Z}{2} (\partial_\mu \sigma)^2 + \frac{Z}{2} (\partial_\mu \pi^a)^2 
    - C_0 \cdot \sigma - \frac{1}{2} \widetilde{m}_0^2 (\sigma^2 + (\pi^a)^2) 
    - \frac{\widetilde{\lambda}_0(\sigma, \pi^a)}{4} (\sigma^2 + (\pi^a)^2)^2  
    \,, 
\end{align}
with 
\begin{align}
    Z &\sim \frac{N_c}{8\pi^2} \ln \frac{\Lambda^2}{{\cal S}^2} 
    \,, \notag \\ 
\widetilde{C}_0 & \sim - (y v_\Phi) \frac{N_c}{\pi^2} \Lambda^2 \,, 
\notag\\ 
    \widetilde{m}^2_0 &\sim \frac{3 N_c}{2 \pi^2} \left[  (y v_\Phi)^2 - 
 (274\,{\rm MeV})^2    \right]
\,, \notag\\ 
 \widetilde{\lambda}_0(\sigma, \pi^a) 
 & \sim  \frac{N_c }{2\pi^2} \ln \frac{\Lambda^2}{(\sigma^2 + (\pi^a)^2)}
\,. \label{Zs}
\end{align}
Here we have focused only on the $(\sigma, \pi^{a=1,2,3}) \approx (\bar{q}q, \bar{q} i \gamma_5 \tau^a q)$ meson sector for the chiral $U(2)_L \times U(2)_R$ partners (because they are the lowest spectra) and have taken the massless quark limit $m=0$, for simplicity. 
This meson-sector Lagrangian can be interpreted as an effective theory 
evaluated at the renormalization scale $\mu = {\cal S} (= \langle \sigma \rangle)$, with 
the compositeness condition: $Z \to 0$ and $\widetilde{\lambda}_0 \to 0$ when $\mu \to \Lambda$. The canonical normalization of the meson fields is achieved by 
the redefinition $\sigma \to \frac{\sigma}{\sqrt{Z}}$ and $\pi^a \to \frac{\pi^a}{\sqrt{Z}}$, so that we have 
\begin{align}
    {\cal L}_{\rm meson} \to  
     \frac{1}{2} (\partial_\mu \sigma)^2 + \frac{1}{2} (\partial_\mu \pi^a)^2 - C\cdot \sigma
    - \frac{1}{2} {m}_0^2 (\sigma^2 + (\pi^a)^2) 
    - \frac{\lambda_0(\sigma, \pi^a)}{4} (\sigma^2 + (\pi^a)^2)^2  
    \,, 
\end{align}
with 
\begin{align}
C & \sim - \frac{2 \sqrt{2}}{\pi} \sqrt{\frac{N_c}{\ln \frac{\Lambda^2}{{\cal S}^2}}} (y v_\Phi)   \Lambda^2 \,, 
\notag\\ 
    {m}^2_0 &\sim \frac{12}{\ln \frac{\Lambda^2}{{\cal S}^2} } \left[  (y v_\Phi)^2 - 
 (274\,{\rm MeV})^2    \right]
\,, \notag\\ 
 {\lambda}_0(\sigma, \pi^a) 
 & \sim  \frac{32 \pi^2}{ N_c  [\ln \frac{\Lambda^2}{{\cal S}^2} ]^2} \ln \frac{\Lambda^2}{(\sigma^2 + (\pi^a)^2)}
\,. \label{meson-Lag}
\end{align}

Based on the meson-sector Lagrangian in Eq.(\ref{meson-Lag}), 
we compute the one-loop effective potential for the 
vacuum expectation value $\langle \sigma \rangle \equiv \sigma_0$ by integrating out 
the meson fluctuations up to one-loop level. 
For the large cutoff $\Lambda \gg \sigma_0$, the effective potential $V_{\sigma_0}^{\rm eff}$ is thus computed as 
\begin{align}
    V_{\sigma_0}^{\rm eff} = C\cdot \sigma_0 + \frac{M^2(\sigma_0)}{2} \sigma_0^2 + \frac{\lambda(\sigma_0)}{4} \sigma_0^4 
    \,, \label{Veff-sigma}
\end{align}
with 
\begin{align}
    M^2(\sigma_0) & \sim m_0^2 + \frac{\lambda_0(\sigma_0,0)}{48 \pi^2} \Lambda^2 
\,, \notag\\ 
\lambda(\sigma_0) & \sim 
\lambda_0(\sigma_0,0) + \frac{3 \lambda_0^2(\sigma_0,0)}{16\pi^2} \ln \frac{\sigma_0^2}{\Lambda^2}
\,. 
\end{align}
%We may identify $\mu = {\cal S} = \sigma_0$ for this effective theory, 
%such that 
The meson loop-corrected potential parameters are further evaluated as 
\begin{align}
    M^2(\sigma_0) & 
    %\sim m_0^2 + \frac{\lambda_0(\sigma_0,0)}{48 \pi^2} \Lambda^2 
    \sim \frac{12}{\ln \frac{\Lambda^2}{{\cal S}^2}} 
    \left[ (y v_\Phi)^2 - (274 \, {\rm MeV})^2 + \frac{\Lambda^2}{18 N_c} \left( \frac{\ln \frac{\Lambda^2}{\sigma^2_0}}{\ln \frac{\Lambda^2}{{\cal S}^2}} \right) \right]
\,, \notag\\ 
\lambda(\sigma_0) & 
%\sim \lambda_0(\sigma_0,0) + \frac{3 \lambda_0^2(\sigma_0,0)}{16\pi^2} \ln \frac{\sigma_0^2}{\Lambda^2} 
\sim 
\frac{32 \pi^2 \ln \frac{\Lambda^2}{\sigma_0^2} }{N_c [\ln \frac{\Lambda^2}{{\cal S}^2}]^2} \left[ 1 - \frac{6}{N_c} \left( \frac{\ln \frac{\Lambda^2}{\sigma^2_0}}{\ln \frac{\Lambda^2}{{\cal S}^2}}   \right)^2 \right]
\,. 
\end{align}
We now find the cancellation condition for the soft-scale breaking term, $M^2(\sigma_0) \equiv 0$, to be 
\begin{align}
  (y v_\Phi) \sim 262 \, {\rm MeV} 
\,,  
\end{align}
or equivalently, 
%so that the chiral phase transition almost goes like 
%the CW-type supercooling. 
\begin{align}
    \frac{y}{\sqrt{\lambda}} \sim  1.3 \times \left( \frac{100\,{\rm MeV}}{\mu_5} \right)  
    \,, \label{y-lambda}
\end{align}
where we have approximated ${\cal S} \sim \sigma_0$ in the log factors.  
As has been discussed in~\cite{Zhang:2025kbu}, more precisely, 
the size of the soft-scale breaking term is constrained to 
realize the CW-type supercooling as 
\begin{align}
    & \Bigg| \frac{M^2(\sigma_0)}{\langle {\sigma_0}\rangle^2} \Bigg|  \lesssim {\cal O}(10^{-10})\,, \notag\\  
    {\rm i.e.,} \qquad 
    & 
    \Bigg| 1 -  \left( \frac{y v_\Phi}{262 \,{\rm MeV}}\right)^2 \Bigg| \lesssim {\cal O}(10^{-11})   
\,,\label{tuning} 
\end{align}
for $\langle \sigma_0  \rangle =$ 400 MeV - 700 MeV.

Another type of enhancement of positive soft-scale breaking would be 
generated from intrinsic $\mu_5$ corrections acting as a catalyzer of 
the chiral/axial breaking under a proper regularization scheme~\cite{Farias:2016let}. 
Including this assist,  the size of $( y\, v_\Phi)$ in Eq.(\ref{tuning}) would be smaller. 
The detailed study to more precisely achieve the ALP-assisted CW-type supercooling is noteworthy 
to be performed in another publication.

\begin{figure}[t]
  \begin{center}
   \includegraphics[width=10cm]{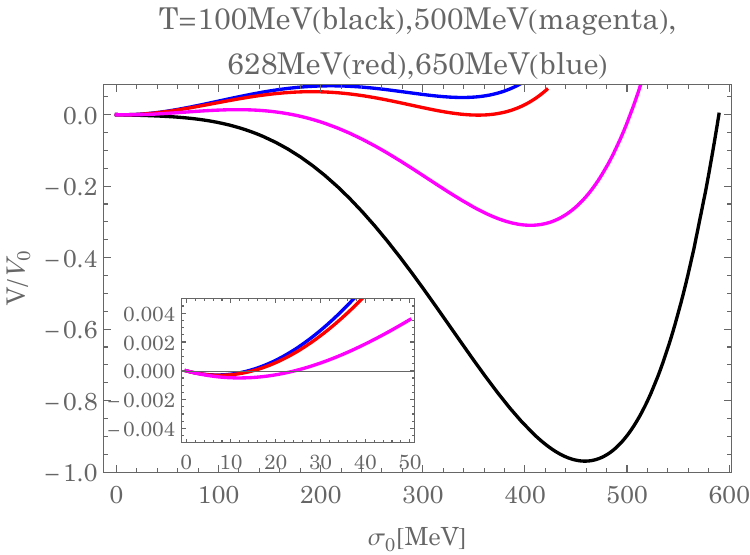}
  \end{center}   
\caption{
The plot of the thermal evolution of the effective potential $V_{\rm eff}$ in Eq.(\ref{Veff-sigma}) with $M^2(\sigma)=0$ plus the Debye mass term correction in Eq.(\ref{T2-term}), normalized by the vacuum energy $V_0 = |V_{\rm eff}(\langle \sigma_0\rangle )|$ measured at the true vacuum $\langle \sigma_0 \rangle_{\rm true} \simeq 461$ MeV. The critical temperature of the first order phase transition is observed as $T_c \simeq 628$ MeV (corresponding to the moment denoted by red curve)
The false vacuum is set away from the origin because of the $\mu_5$-induced tadpole term, which is at $\langle \sigma_0 \rangle_{\rm false} \simeq 7.9$ MeV when $T=T_c$, as depicted in the zoomed-in figure on the bottom-left interior panel.  
}  
\label{Veft-supercooling-plot}
\end{figure}

Simply adding to the effective potential in Eq.(\ref{Veff-sigma}), with the $M^2=0$ condition in Eq.(\ref{y-lambda}),  
the Debye mass terms arising from the quark (in Eq.(\ref{high-T-exp})) and meson thermal fluctuations 
\begin{align}
    V_{T^2} = \left( \frac{N_c}{3} T^2 + \frac{\lambda_0(\sigma_0, 0)}{4} T^2 \right) \cdot \frac{\sigma_0^2}{2}   
\,, \label{T2-term}
\end{align}
we can roughly evaluate the thermal evolution of the total potential around the criticality. 
This rough prescription can work fine because it turns out that $T_c \gg \sigma_0$ around the true vacuum, so that higher orders in $(\sigma_0/T)$, including thermal cubic and CW-type log terms, can be subdominant.   
See Fig.~\ref{Veft-supercooling-plot}, where 
the critical temperature is observed around $T \simeq 628$ MeV, at which temperature 
the true vacuum (at $\langle \sigma_0 \rangle \simeq 461$ MeV) degenerates with 
the false vacuum (at $\langle \sigma_0 \rangle \simeq 7.9$ MeV) created by the tadpole term $\propto \mu_5$. 
Note that the $\mu_5$-induced tadpole term is sizable, $C \simeq - (541\, {\rm MeV})^3$, which is much greater in magnitude than the current-quark mass-induced tadpole of ${\cal O}(m\cdot \Lambda^2)={\cal O}(100\,{\rm MeV})$. 
%This fact is also reflected in the larger $\langle \sigma_0\rangle $, $\langle \sigma_0 %\rangle \simeq 463$ MeV. 
As has been noted in the beginning of the present section (and also see footnote~\ref{foot}), 
all the $\mu_5$ term contributions will be dropped out when the QCD sphaleron gets confined and frozen out. 
In the present scenario, the confining phase is realized after passing through the supercooling, 
when $T=T_n \ll T_c \simeq 628$ MeV, which is expected to be at $T = {\cal O}(100 \,{\rm MeV})$ just like the ordinary QCD scenario.

The false vacuum still thermally shifts from $\langle \sigma_0 \rangle_{\rm false} \simeq 7.9$ MeV at $T=T_c$ to a larger value as $T$ cools. 
At $T = T_n (\ll T_c) \sim {\cal O}(100)$ MeV, the potential barrier is gone and 
the false vacuum is no longer an inflection point (i.e., gets unstable).  
At this moment, the $\sigma$ field starts to roll from the false vacuum down to the true vacuum at $\langle \sigma_0 \rangle_{\rm true} \simeq 461$ MeV. This is a classical roll, not a quantum tunneling. The operated mechanism there is precisely the same as the dynamical realization of the CW-type small-field inflation~\cite{Zhang:2023acu,Liu:2024xrh}. 
This classical roll includes several steps: a slow roll, which could cause a mini-inflation; a fast roll, which could be relevant to nonadiabatic particle and entropy productions; and an oscillation around the true vacuum. 
Thus, this potentially involves rich cosmology. GW and/or PHB production could be realized~\cite{Zhang:2024vpp,Zhang:2025kbu} during the slow-roll (mini-inflation) and/or nonadiabatic fast roll (preheating, just like the QCD preheating~\cite{Wang:2022ygk,Wang:2024tda}), and/or oscillation (reheating).  
These intriguing issues deserve to be addressed in detail in another publication.

\subsection{ALP phenomenology in present-day-universe} 

Passing through the reheating process, the Universe cools down to  
sufficiently low temperatures much below the confinement scale of ${\cal O}(100\,{\rm MeV})$ (or the expected nucleation at $T_n$). 
At this point, the QCD sphaleron has already been decoupled because of the 
Boltzmann suppression for the sphaleron rate, so have been all the $\mu_5$ contributions in the effective potential, so that the large negative mass squared, i.e., the conventional dynamical symmetry breaking has been recovered. 
The meson loop corrections are then subleading -- since even the pions act as heavy enough due to the negative mass squared -- as also noted in the previous subsection, so the theory is settled back in the chiral broken phase simply following 
the original NJL-MF dynamics with the vacuum $\langle {\cal S} \rangle \simeq 394$ MeV. 
This is the vacuum characterizing the hadron phase in the present-day Universe. 
However, it turns out that the present-day Universe includes not only the ordinary QCD hadrons, but also light ALPs.

Although the $\mu_5$ terms are gone, 
the $\Phi$-ALP field still gets a mass squared correction 
from the quark loop, enough to develop its vacuum expectation value in the 
confining and chiral broken phase. 
Noting that the $\Phi$-ALP field does not couple to the meson sector 
up to the one-loop level, we focus only on the quarkonic sector to evaluate 
the development of the $\phi$ potential at this point.  
That is the thermodynamic potential form in Eq.(\ref{Phi-shift}) with $\langle {\cal S}  \rangle$ being replaced by $\langle {\cal S}\rangle \simeq 394$ MeV, 
which indeed provides an effective potential for $\phi = (y \phi_s)$ as follows:  
\begin{align}
    \Omega(\phi = y \phi_s)\Bigg|_{T \ll T_c} 
    \sim 
    -  \frac{N_c N_f}{4 \pi^2} \Lambda^2 (\langle {\cal S}\rangle - \phi)^2 + \frac{N_f N_c}{16 \pi^2} (\langle {\cal S}\rangle - \phi)^4 \ln \frac{\Lambda^2}{(\langle {\cal S}\rangle - \phi)^2}  
    \,.  
\end{align}
%where terms of ${\cal O}({\cal S}^2/(4\pi^2))$ have been dropped since ${\cal S}^2 \ll \Lambda^2$. 
This, together with the quartic coupling term $\lambda  |\Phi|^4 = (\lambda/y^4) |\phi|^4$, 
yields the vacuum expectation value of $\Phi$, $(v_\Phi)_0$, 
\begin{align}
    (v_\Phi)_0 \approx - \left( \frac{y^4}{\lambda} \frac{N_cN_f}{2\pi^2} \langle {\cal S}\rangle \Lambda^2 \right)^{1/3}
    \,, 
\end{align}
for $(v_{\Phi})_0 \ll \langle {\cal S}\rangle$.  
Here we have assumed $y^4\ll \lambda$, so that the $\lambda$-quartic term highly dominates over the loop-induced quartic term. 
This $(v_{\Phi})_0 $ is generated essentially by the explicit breaking, via the $y$-Yukawa coupling term in Eq.(\ref{yS}) with $\langle \bar{q}q \rangle \neq 0$.

Around the vacuum at $(v_\Phi)_0$, 
the $\Phi$-ALP field thus fluctuates to develop the mass along its radial component $\chi$ in the polar decomposition form, 
\begin{align} 
\Phi= \chi \cdot e^{i a/(v_\Phi)_0}
\,. \label{polar}
\end{align} 
The $\chi$-ALP mass $m_{\chi}$ is then evaluated as 
\begin{align}
    m_{\chi} \approx 
    \frac{\sqrt{\lambda}}{y} |(v_\Phi)_0| 
    &\approx \left( \frac{\lambda^{1/2}}{y} \frac{N_cN_f}{2 \pi^2} \langle {\cal S}\rangle \Lambda^2 \right)^{1/3} 
    \notag\\ 
    &\simeq 346\,{\rm MeV} \times \left( \frac{\lambda^{1/2}}{y} \right)^{1/3} 
    \simeq 318 \, {\rm MeV} \times \left(\frac{\mu_5}{100\,{\rm MeV}} \right)^{1/3} 
    \,, \label{mchi}
\end{align}
for $N_c=3$ and $N_f=2$. Here we have ignored the quark-loop induced-wavefunction renormalization correction for $\Phi$, and in the last equality, we have used the parameter condition in Eq.(\ref{y-lambda}). 
The Nambu-Goldstone boson mode $a$ in $\Phi$ also gets the mass and axionlike potential due to the generation of the explicit-breaking tadpole induced by $\langle \bar{q}q \rangle \neq 0$ in Eq.(\ref{yS}): 
\begin{align}
    V(a) = y (- v_{\Phi})_0 \langle - \bar{q}q \rangle \cos \frac{a}{(v_\Phi)_0} 
    \,. 
\end{align} 
From this potential, the $a$-ALP mass reads 
\begin{align}
    m_a^2 & = \frac{y \langle - \bar{q}q  \rangle}{(-v_\Phi)_0} 
\notag\\ 
& \simeq  \left[ 196 \, {\rm MeV} \times \left( \frac{\lambda}{y} \right)^{1/6}\right]^2
\simeq \left[ 26 \, {\rm MeV} \times \left(\frac{y}{10^{-5}} \right)^{1/6} \cdot \left( \frac{\mu_5}{100\,{\rm MeV}} \right)^{1/3} \right]^2 
\,.\label{ma}
\end{align}
%where we have used the MFA $S \equiv \langle \bar{q}q \rangle$  
Note that $m_a < m_\chi$ for any of $\lambda$ and $y$, as long as $y \ll 1$.

The ALP-Yukawa coupling to light quarks as in Eq.(\ref{yS}) has severely been constrained via the ALP decay to diphoton 
by the astrophysical, cosmological, and collider experimental observations.  
See the public website at {\url{ https://cajohare.github.io/AxionLimits}} or {\url{ https://cajohare.github.io/AxionLimits/docs/ap.html}}. 
The diphoton coupling for the $\Phi$-ALP field is generated  
by mixing with the QCD-$\sigma$ (for the $\chi$-ALP) and $-\eta'$ mesons (for the $a$-ALP), arising through the Yukawa coupling $y$ in Eq.(\ref{yS}). 
We may roughly evaluate the $\chi(a)-\gamma-\gamma$ coupling at the $\chi/a$ onshell,  $g_{\chi(a)\gamma\gamma}$, as 
\begin{align}
   g_{\chi(a)\gamma\gamma} \approx   
   \frac{y\, \Lambda^2}{\sqrt{(m_{\chi/a}^2 - m_{f_0(500)/\eta'}^2)^2 + m_{f_0(500)/\eta'}^2 \Gamma^2_{f_0(500)/\eta'}}} \times g_{f_0(500)/\eta' \gamma\gamma} 
\,,  \label{reso}
\end{align}
where $\Gamma_{f_0(500)/\eta'}$ denotes the total widths of $f_0(500)$ and $\eta'$, respectively. In Eq.(\ref{reso}) the denominator form comes from the $f_0(500)\eta'$ propagator in the Breit-Wigner form, evaluated at the transfer momentum squared $q^2=m_{\chi/a}^2$, and the model-dependent factor $\sim y \Lambda^2$ denotes the 
mass mixing strength between the ALPs and those mesons. 
In the present evaluation, the latter part has been derived from 
the NJL-MFA based on the effective potential form in Eq.(\ref{Phi-shift}), with 
${\cal S}$ and $v_\Phi$ replaced by $\eta'$ and $(v_\Phi)_0 \sin a/(v_\Phi)_0$ following the polar decomposition of $\Phi$ in Eq.(\ref{polar}) applied to the $y$-Yukawa term in Eq.(\ref{yS}). 
The wavefunction renormalization of $f_0(500)/\eta'$  has also been taken into account, which in the NJL-MFA yields $\sim \frac{N_c}{8 \pi^2} \ln \frac{\Lambda^2}{\langle {\cal S}\rangle^2}$ as in Eq.(\ref{Zs}) for 
the inverse propagator in the chiral broken phase with $(v_\Phi)_0 \ll \langle {\cal S}\rangle$. 
%Although the present NJL model only includes two flavors, the essential cutoff $\Lambda$ %dependence on the mixing strength and wavefunction-renormalization factor should be generic up 5to a factor of ${\cal O}(1)$, 
%for the flavor singlet mesons, such as $\eta_0$ for the two-flavor case and $\eta'_0$ for the three-flavor case.  
%Therefore, 
The thus estimated net model-dependent factor $\sim y \Lambda^2$ in Eq.(\ref{reso}) covers the sufficient size in order of magnitude.

%Note that both the $\chi$ and $a$-ALP masses in Eqs.(\ref{mchi}) and (\ref{ma}) 
%are predicted to be so large because $y \ll 1$, which grow like $\propto y^{-1/3}$ and 
%$y^{-1/6}$, respectively. 
The $f_0(500)$ coupling to diphoton has not yet been clearly determined in experiments~\cite{ParticleDataGroup:2024cfk}. 
Moreover, the $\chi$ resonance will be contaminated with the $f_0(500)$ production, e.g., in the $\pi\pi$ scattering process. This production channel will thus be challenging for the probe of the present ALP-assisted model. 
Therefore, at this point, we disregard discussing the phenomenological constraint and consequence of $\chi$.

The total width of $\eta'$ is available from the Particle Data Group~\cite{ParticleDataGroup:2024cfk}; at the central value, $\Gamma_{\eta'} \simeq 0.188$ MeV at $m_{\eta'} \simeq 958$ MeV. We neglect the small mixing between 
$\eta'$ and $\eta$. 
Regarding the $\eta'$ mixing-induced ALP-photon coupling, in that case, 
we may refer to the case of composite ALP models at around the QCD scale~\cite{Wang:2024vlm}, 
where $g_{\eta \gamma\gamma} \sim 10^{-3}\,{\rm GeV}^{-1}$, which corresponds to 
the soft-${\eta'}$ limit value based on a chiral effective model approach. 
%Since $y \lesssim 1$, hence $m_a \gtrsim m_{\eta'}$, 
%we approximate 

\begin{figure}[t]
  \begin{center}
   \includegraphics[width=10cm]{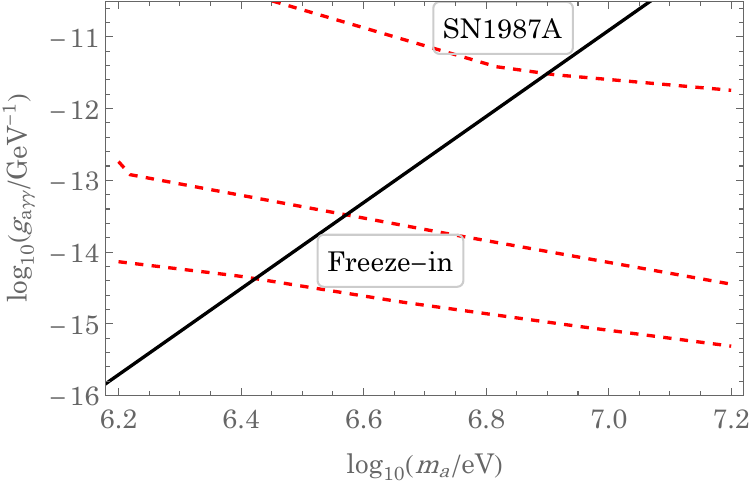}
  \end{center}   
\caption{
The predicted curve of $g_{a\gamma\gamma}$ as a function of $m_a$ with the soft-scale breaking-cancellation condition in Eq.(\ref{y-lambda}) taken into account. 
The plot range 
%has taken into account the lower bound on $m_a$, set when $y=\lambda$ to 
ensures the perturbatively small enough quark loop correction to the ALP sector (i.e., 
$y^4 \ll \lambda$, or equivalently $m_a \ll 169$ MeV). 
%The spike structure is generated at the resonance point when $m_a = m_{\eta'}$, based %on the resonant form factor structure of $g_{a\gamma\gamma}$ in Eq.(\ref{agg-focus}). 
All the domains plotted not overlapping with the constraint regimes survive over the current experimental constraint summarized in the website~{\url{ https://cajohare.github.io/AxionLimits}} or {\url{ https://cajohare.github.io/AxionLimits/docs/ap.html}}. The case with $m_a< 1$ MeV has also been ruled out by the bound from the ``Cosmic Background". }  
\label{g-ma-plot}
\end{figure}

Thus, we focus on the $a-\gamma-\gamma$ coupling, 
\begin{align}
    g_{a\gamma\gamma} &\approx   
   \frac{y\, \Lambda^2}{\sqrt{(m_{a}^2 - m_{\eta'}^2)^2 + m_{\eta'}^2 \Gamma^2_{\eta'}}} \times g_{\eta' \gamma\gamma} 
 %   \notag\\ 
 %   & \sim y \times \frac{|(v_\Phi)_0|}{f_\pi} \times  10^{-3} \, {\rm GeV}^{-1}
    \,. \label{agg-focus}
\end{align}
Solving Eq.(\ref{ma}) with respect to $y$; $y \simeq  10^{-5} \times (m_a/26\,{\rm MeV})^6 $ for $\mu_5 =100$ MeV, 
and putting this $y$ into Eq.(\ref{agg-focus}), we numerically write $g_{a\gamma\gamma}$ as a function of $m_a$. 
This function draws the theoretical prediction curve on the plane 
$(m_a, g_{a\gamma\gamma})$ available at the AxionLimits ({\url{ https://cajohare.github.io/AxionLimits}} or {\url{ https://cajohare.github.io/AxionLimits/docs/ap.html}}). 
The predicted curve on $(m_a, g_{a\gamma\gamma})$ is plotted in Fig.~\ref{g-ma-plot}, with the soft-scale breaking-cancellation condition in Eq.(\ref{y-lambda}) and the perturbativity $y^4 \lambda$ taken into account (i.e., 
$y^4 \ll \lambda$, or equivalently $m_a \ll 169$ MeV). 
The $\mu_5$ sensitivity on $g_{a\gamma\gamma}$ is only quadratic: $g_{\rm a\gamma\gamma}\propto \mu^2_5$, so the predictive curve in Fig.~\ref{g-ma-plot} does not substantially change as long as $\mu_5$ is around the typical QCD scale, in a range of 100 - 200 MeV.

The $a$-ALP cannot be heavier than $\sim 169$ MeV so as to keep  
the perturbatively small enough quark loop correction to the ALP sector (i.e., 
$y^4 \ll \lambda$). 
Furthermore, the current bound severely constrains $m_a$  to be $\sim 4$ MeV - 8 MeV or $\sim 1$ - 2.5 MeV, which has been reflected in the figure. 
The case with $m_a< 1$ MeV has also been ruled out by the bound from the ``Cosmic Background". 
%the curve can be consistent with the current experiments, astrophysical, and cosmological observations. However, this lower bound on $m_a$ gives too large $y$: $y \sim \lambda$ in magnitude, therefore, the actual lower bound is set when $y=\lambda$, i.e., $m_a \simeq 124$ MeV. Figure~\ref{g-ma-plot} has already taken into account this limit. 
As a benchmark, 
we have $g_{a\gamma\gamma} \sim 10^{-12}\,{\rm GeV}^{-1}$ for $m_a \sim 5$ MeV, which corresponds to the case with $y \sim 5 \times  10^{-10}$ with $\lambda \sim 10^{-19}$.

This heavy ALP is unlikely to be probed or excluded by the upcoming prospect 
experiments, such as the Electron Ion Collider experiment~\cite{Balkin:2023gya} or 
the Search for Hidden Particles (SHiP, at CERN) experiment and the Beam Dump eXperiment (BDX, at JLab)~\cite{Patrone:2025fwk}. 
However, indirect evidence could be tested via the GW and PBH productions sourced from the CW-type supercooling chiral phase transition, as has been suspected in the previous section.

\section{Conclusion}  
\label{concl}

In conclusion, 
we have clarified that the chiral phase transition in QCD-like theories can be CW type supercooling. 
This possibility paves the way to a new search window for 
new physics to be probed by GW and PBH productions. 
We have discussed this possibility by monitoring ordinary QCD setup in a view of 
a two-flavor NJL model in a hot and/or dense medium, at the MFA level. 
It turned out that the identification essentially follows the scale violation classification: soft-scale breaking and quantum scale anomaly of the CW-type. 
The CW-type chiral-phase transition is possible to realize when 
ordinary QCD is coupled to or replaced by Beyond the Standard Model in a nontrivial way. 
This is what is called the ALP-assisted CW-type supercooled chiral phase transition. 
This new scenario predicts rich cosmological and phenomenological consequences around the QCD scale: a post supercooled CW-type small-field (mini-) inflation; preheating; reheating along with GW and/or PBH production.

The proposed scenario predicts a heavy ALP with mass $\sim $ 5 MeV in the present-day Universe. 
Since the couplings $y$ and $\lambda$ are tiny, this ALP could also resolve the strong CP problem in a way similar to the conventional axion solution, consistently with 
the current bound from the neutron electric dipole moment. 
The natural inflation of this ALP can also be realized in the earlier epoch of the thermal history. The cosmological evolution from the inflationary to QCD epochs would also be worth pursuing.

Arguments similar to the present NJL description can also be applied to bosonic models, e.g., quark meson models and linear sigma models, or more generic boson-fermion system. The key ingredient is at any rate identifying the CW-type scale anomaly and the size of the soft-scale breaking. 
Such a more general investigation would also be noteworthy.

In closing, the CW-type chiral phase transition is realized because of the dominance of 
the quantum scale anomaly of the logarithmic form around the chiral criticality. 
This qualitative feature would still be operative even including gluonic or alternatively non-local (momentum dependent) interactions among quarks, because of persistence of the logarithmic renormalization group running for the potential (more precisely 
the induced quartic coupling of the CW scalar, which is identified as the bosonized quark bilinear in QCD-like theories). 
%In fact, it has been indicated that gluonic contributions tend to merely quantitatively %affect the chiral phase transition nature~\cite{}. 
This characteristic logarithmic scale-anomaly is seen in any regularization scheme~\cite{Kohyama:2015hix}, 
unless dimensional reduction takes place as in the case with a strong magnetic field.  
Thus, our present proposal would be free from the intrinsic regularization 
dependence on the quantitative deformation of the chiral phase structure as has been addressed in the literature~\cite{Kohyama:2015hix}.

\section*{Acknowledgments} 
This work was supported in part by the National Science Foundation of China (NSFC) under Grant No.11747308, 11975108, 12047569, 
and the Seeds Funding of Jilin University (S.M.). 
The work by M.K. is supported by RFIS-NSFC under Grant No. W2433019.
The work of A.T. was partially supported by JSPS  KAKENHI Grant Numbers 20K14479, 22K03539, 22H05112, and 22H05111, and MEXT as ``Program for Promoting Researches on the Supercomputer Fugaku'' (Simulation for basic science: approaching the new quantum era; Grant Number JPMXP1020230411, and 
Search for physics beyond the standard model using large-scale lattice QCD simulation and development of AI technology toward next-generation lattice QCD; Grant Number JPMXP1020230409).

\end{document}